\documentclass[epj,final]{svjour}
\usepackage{latexsym,amsmath}
\usepackage{graphics}
\usepackage{graphicx}
\usepackage{amssymb}
\usepackage[latin1]{inputenc}
\begin{document}

\title{Critical-state effects on microwave losses in type-II superconductors}
\author{M. Bonura,  E. Di Gennaro, A. Agliolo Gallitto, and M. Li Vigni}
\institute{CNISM and Dipartimento di Scienze Fisiche ed
Astronomiche, Università di Palermo, Via Archirafi 36, I-90123
Palermo, Italy }

%\maketitle

\abstract {We discuss the microwave energy losses in
superconductors in the critical state. The field-induced
variations of the surface resistance are determined, in the
framework of the Coffey and Clem model, by taking into account the
distribution of the vortex magnetic field inside the sample. It is
shown that the effects of the critical state cannot generally be
disregarded to account for the experimental data. Results obtained
in bulk niobium at low temperatures are quantitatively justified.
}
\authorrunning{M. Bonura et \textit{al.}}
\titlerunning{Critical-state effects on microwave losses}
\maketitle

\section{Introduction}
Investigation of fluxon dynamics in type-II superconductors is of
great interest for both fundamental and applicative aspects. From
the basic point of view, it allows measuring the relative
magnitude of elastic and viscous forces, which rule the motion
regime of the fluxon lattice \cite{gittle,golo,TALVA,noi}. From
the technological point of view, it allows determining the
critical current, accounting for the energy losses and
investigating the presence of irreversible phenomena, which are
important factors for the implementation of superconductor-based
devices \cite{libromw}.

A suitable method to investigate the fluxon dynamics consists in
determining the field-induced energy losses at microwave (mw)
frequencies, by measuring the surface resistance, $R_s$
\cite{golo}. In the absence of static magnetic fields, the
variation with the temperature of the condensed-fluid density
determines the temperature dependence of $R_s$. On the other hand,
the field dependence of $R_s$ in superconductors in the mixed
state is determined by the presence of fluxons, which bring about
normal fluid in their cores, as well as the fluxon motion
\cite{gittle,golo,TALVA,noi,CC4,BRANDT,noiPRB}.

Coffey and Clem (CC) have elaborated a comprehensive theory for
the electromagnetic response of type-II superconductors in the
mixed state \cite{CC4}, taking into account flux creep, flux flow,
and pinning effects in the framework of the two-fluid model. The
theory has been developed with two basic assumptions: i)
inter-vortex spacing much less than the field penetration depth;
ii) uniform vortex distribution in the sample. With these
assumptions, the local vortex magnetic field,
$B(\textit{\textbf{r}})$, averaged over distances larger than
several inter-vortex spacings, is spatially uniform. As pointed
out by the authors \cite{CC4}, the first assumption is valid when
the external field, $H_0$, is greater than $2H_{c1}$; whereas, the
second one does not take into account the magnetic history of the
sample.

It is well known that the \textit{H - T} phase diagram of type-II
superconductors is characterized by the presence of the
irreversibility line, $H_{irr}(T)$, below which the magnetic
properties of the superconducting sample become irreversible. The
application of a DC magnetic field smaller than $H_{irr}(T)$
develops a critical state of the vortex lattice \cite{BEAN,KIM}
and the assumption that $B$ is uniform over the sample is not
longer valid. Although the critical state is a metastable energy
state, the relaxation times toward the thermal equilibrium state
can be much longer than the time during which the measurements are
performed \cite{yesherun}. In this case, the effects of the
critical state can be revealed in all the superconducting
properties involving the presence of fluxons in the sample.

When the sample is in the critical state, the CC model does not
account for the experimental results; for instance, no magnetic
hysteresis in the $R_s(H_{0})$ curves is expected. Though
different authors have justified the hysteretic behavior of the
$R_s(H_{0})$ curves by considering critical-state effects in the
fluxon lattice \cite{TINKHAM,SRIDHAR}, the field dependence of the
surface resistance for non-uniform flux profile has never been
quantitatively investigated.

In this paper, we show that, when the superconducting sample is in
the critical state, the CC theory has to be generalized by
properly taking into account the flux distribution inside the
sample. We suggest a way to considering the critical-state
effects. We will show that the $R_s(H_0)$ curve strongly depends
on the specific profile of the magnetic induction $B$, determined
by the field dependence of the critical current density $J_c(B)$.
We will report experimental results of the field-induced
variations of $R_s$ in a niobium bulk sample in the critical
state; the results are quite well accounted for by considering a
specific $B$ profile inside the sample.

\section{The model}

In the London local limit, the surface impedance is proportional
to the complex penetration depth $\widetilde{\lambda}$ of the
\textit{em} field. In particular,
\begin{equation}\label{Rs}
    R_s=-\mu_{0}\omega ~\mathrm{Im}[{\widetilde{\lambda}(\omega,B,T)}].
\end{equation}
In the CC model, $\widetilde{\lambda}(\omega,B,T)$ is given by
\begin{equation}\label{lambdat}
    \widetilde{\lambda}(\omega,B,T)=\sqrt{\frac{\lambda^{2}(B,T)+
    (i/2)\widetilde{\delta}_{v}^{2}(\omega,B,T)}
    {1-2i\lambda^{2}(B,T)/\delta _{nf}^{2}(\omega,B,T)}},
\end{equation}
with
\begin{equation}\label{lamda0}
\lambda(B,T) = \frac{\lambda_0}{\sqrt{[1-(T/T_c)^4][1- B
/B_{c2}(T)]}},
\end{equation}
\begin{equation}\label{delta0}
\delta_{nf}(\omega,B,T) = \frac{\delta_0}{\sqrt{1-[1-(T/T_c)^4][1-
B /B_{c2}(T)]}},
\end{equation}
where $\lambda_0$ is the London penetration depth at  $T = 0$ and
$\delta_0$ is the normal-fluid skin depth at $T = T_c$.\\
$\widetilde{\delta} _{v}$ is the effective complex penetration
depth arising from the vortex motion and depends on the relative
magnitude of the viscous-drag and restoring-pinning forces. An
important parameter, which determines the regime of the fluxon
motion, is the so-called depinning frequency, $\omega_c$, defined
by the ratio between the restoring-force and viscous-drag
coefficients.  When the frequency of the $em$ wave, $\omega$, is
much larger than $\omega_c$ the fluxon motion is ruled by the
viscous-drag force and the induced $em$ current makes fluxons move
in the flux-flow regime, even for $H_0<H_{irr}(T)$. On the
contrary, for $\omega \ll \omega_c$ the motion of fluxons is ruled
by the restoring-pinning force. In the following, we will devote
the attention to the case in which the vortex lattice moves in the
flux-flow regime, where the field-induced energy losses are
noticeable. In this case, considering the expression of the
viscous coefficient proposed by Bardeen and Stephen \cite {BS}, it
results $\widetilde{\delta}_v^2=\delta_0^2B/B_{c2}(T)$. The real
part of $\widetilde{\lambda}$ defines the AC penetration depth,
$\lambda_{ac}$.

Since the mw losses depend on the local magnetic field
$B(\textit{\textbf{r}})$, when it is not uniform over the sample,
different regions of the sample contribute to the energy losses
differently; this should occur when the sample is in the critical
state. Generally, the critical state develops at temperatures
smaller enough than $T_c$, where the pinning effects are
significant; in this case, the energy losses are mainly related to
the motion of fluxons induced by the mw current. So, a
discriminating criterium, to determine in what extent the
non-uniform $B$ profile affects the energy losses, consists in
evaluating the variation of the DC magnetic induction in the
region where fluxons feel the Lorentz force.

\begin{figure}[h]
\centering
\includegraphics[width=7cm]{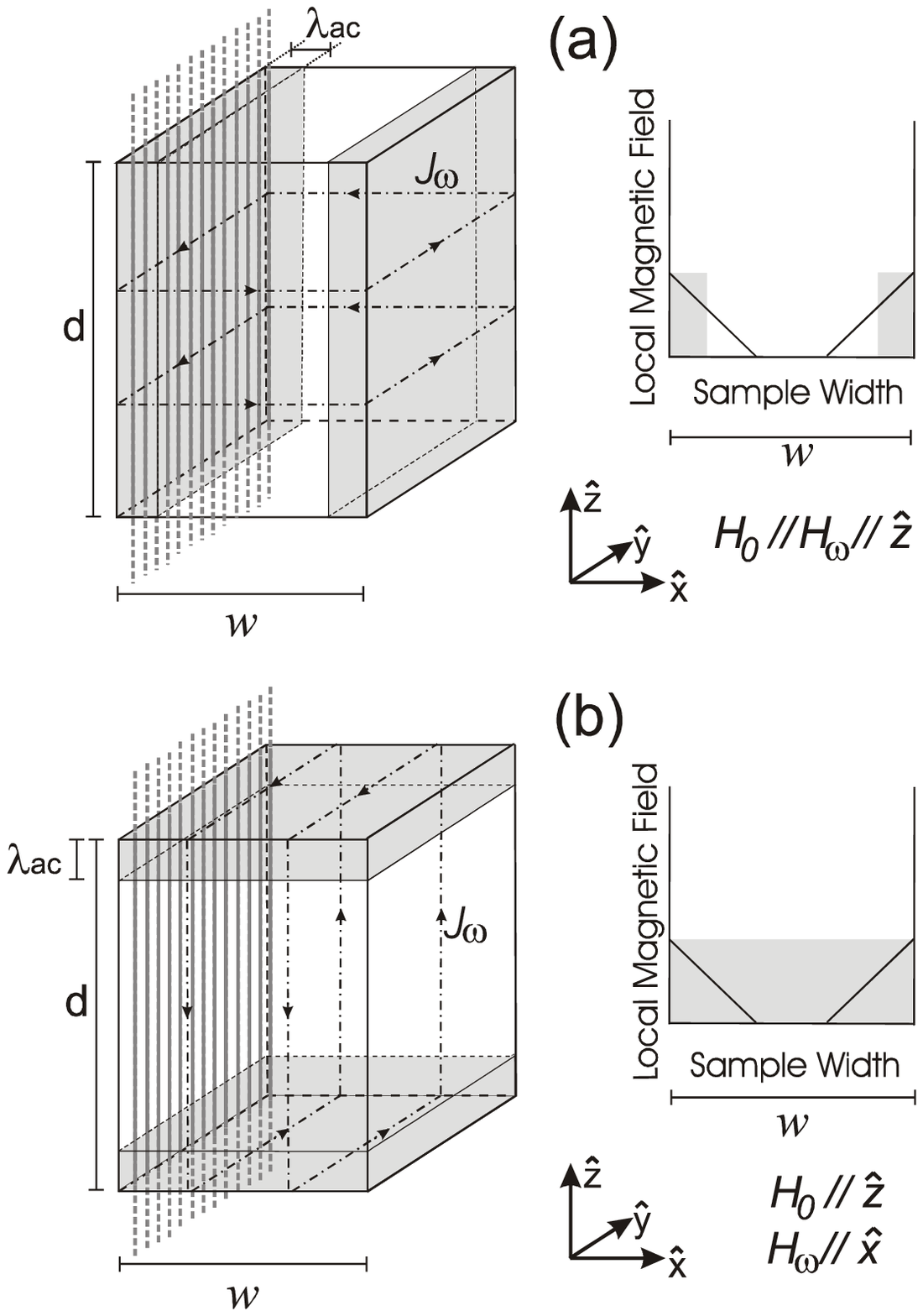}
\caption{Distribution of the fields and current in two different
geometries: $\textit{\textbf{H}}_\omega\|\textit{\textbf{H}}_0~
(a)$; $\textit{\textbf{H}}_\omega\bot \textit{\textbf{H}}_0~(b)$;
the shadowed area indicates the regions of the sample interested
by the vortex motion.} \label{f1}
\end{figure}

Fig.~\ref{f1} describes the distribution of the fields and the mw
current in the two geometries:
$\textit{\textbf{H}}_\omega\|\textit{\textbf{H}}_0$ (a);
$\textit{\textbf{H}}_\omega\bot \textit{\textbf{H}}_0$ (b). For
the sake of clearness, we limit the analysis to a slab geometry
(sample width $w$ and sample height $d$, with $\lambda_{ac}\ll w,
d$). We suppose that the static magnetic field, $H_0$, is greater
than $H_{c1}$ and that the sample is in a critical state à la
Bean, i.e. characterized by a field-independent $J_c$ \cite{BEAN};
however, the analysis applies to any $J_c(B)$ dependence. When the
mw magnetic field is parallel to the $z$-axis, the mw current
penetrates in the surface layers of the sample, of width
$\lambda_{ac}$, in the \textit{x - y} plane. If
$\textit{\textbf{H}}_\omega\|\textit{\textbf{H}}_0$, only the
fluxons in these layers experience the Lorentz force, due to the
mw current, and the fluxon lattice undergoes a compressional
motion. In Fig.~1 (a) we analyze the regions indicated by the
shadowed area. In this case, if the variation of $B$ in these
regions is negligible, i.e. $J_c\lambda_{ac}\ll\ H_0$, the vortex
magnetic field can be considered as uniform. A different situation
occurs when $\textit{\textbf{H}}_{\omega}\bot
\textit{\textbf{H}}_{0}$, as shown in Fig.~\ref{f1}~(b); in this
case, all the vortices present in the sample are involved in the
motion, the fluxon lattice undergoes a tilt motion \cite{BRANDT},
$\lambda_{ac}$ determines the part of the flux line in which the
Lorentz force acts; the approximation of $B$ uniform would be
proper only if the sample width is such that $J_cw\ll\ H_0$. So,
for bulk samples, it is expected that the specific profile of $B$
strongly affects the $R_s(H_0)$ curve; indeed, different regions
of the sample contribute in different extent to the mw energy
losses. In the following, we will refer to this field geometry,
where the effects of the non-uniform profile of $B$ most affect
the $R_s(H_0)$ curve.

Brandt \cite{BRANDT} has shown that the compressional and tilt
motion of the fluxon lattice can be described by the same
formalism and the same AC penetration depth results because the
compression and tilt moduli are approximately equal. So, the
results obtained from the CC model are valid even in the case of
tilt motion provided that $B$ is uniform over the region where the
Lorentz force is active.

When the DC magnetic induction is not uniform, one can subdivide
the sample surface in different regions, in each of them
$B(\textit{\textbf{r}})$ can be considered uniform. Each part of
the surface is characterized by a different $R_{s}$ value, due to
the local magnetic induction, and the energy losses of the whole
sample are determined by the surface-resistance contributions of
each region. The measured surface resistance is an average over
the whole sample:
\begin{equation}\label{Rs}
    R_{s}= \frac{1}{S}\int_\Sigma R_s(|B(\textit{\textbf{r}})|)\,
    dS\,,
\end{equation}
where $\Sigma$ is the sample surface, $S$ is its area and
\textit{\textbf{r}} identifies the surface element; the main
contribution comes from the sample regions where
$\textit{\textbf{H}}_0 \times \textbf{\textit{J}}_\omega \neq 0$.

Usually, in order to disregard the geometrical factor of the
sample, the reported experimental results of the surface
resistance are normalized to the normal-state value at $T=T_c$,
$R_n$. In the flux-flow regime, in order to calculate $R_s/R_n$,
by Eqs.~(1--4), it is necessary to know the ratio
$\lambda_0/\delta_0$ and the upper critical field. Moreover, to
take into account the critical-state effects, by Eq.~(\ref{Rs}),
it is also necessary to know the $B$ profile inside the sample,
determined by $J_c(B)$. We would remark that the analysis can be
generalized, in the different motion regimes, by introducing the
field dependence of the depinning frequency in
$\widetilde{\delta}_v (\omega, B, T)$; however, this introduces
further fitting parameters.

\begin{figure}[h]
\centering
\includegraphics[width=7cm]{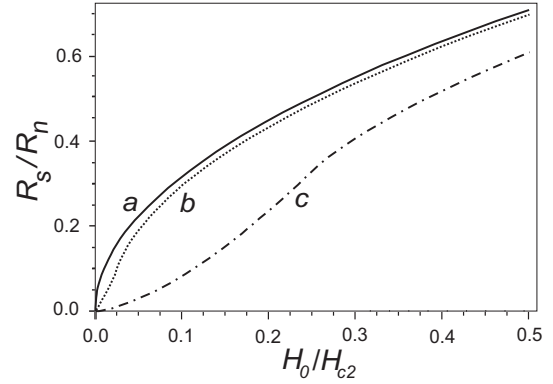}
\caption{Normalized $R_s(H_{0})$ curves, expected in the flux-flow
regime, in the different cases: $(a)$~$B$ uniform; $(b)$ critical
state à la Bean with $H^{*}=H_{c2}/40$ and $(c)$ $H^{*}=H_{c2}/4$.
$T=T_c/4$, $\lambda_0/\delta_0=10^{-2}$.} \label{f2}
\end{figure}

Fig.~\ref{f2} shows $R_s/R_n$ expected in the flux-flow regime as
a function of the reduced field, $H_0/H_{c2}$, in three different
cases. Curve $(a)$ is the one expected from the CC model
($J_c=0$). Curves $(b)$ and $(c)$ have been obtained with
$J_c=J_{c0}$, independent of $B$, and two different values of the
full penetration field $H^{*}$ \cite{BEAN}. All the curves have
been obtained with $T=T_c/4$ and $\lambda_0/\delta_0=10^{-2}$;
furthermore, for simplicity, it was supposed $H_{c1}=0$. It is
evident that, taking into account the distribution of $B$,
different field dependencies of $R_s$ arise. In particular, a
comparison between curves $(b)$ and $(c)$ shows that the greater
$H^{*}$, the lower the value of $R_s$; this occurs because the
induction field of the whole sample decreases on increasing $H^*$.

Another important characteristic of the $R_s(H_0)$ curves is the
change of concavity occurring at $H_0=H^*$. When the fluxon
lattice moves in the flux-flow regime and $B$ is uniform, it is
expected $R_s\propto\sqrt B$, with $B\approx\mu_0H_0$, which leads
to a negative concavity of the $R_s(H_0)$ curve. For the general
case of spatially-dependent flux density, the shape of the
$R_s(H_0)$ curve is determined by the external-field dependence of
$B$. In particular, when in the whole sample the local magnetic
induction linearly depends on the external field a negative
concavity of the $R_s(H_0)$ curve is expected. This occurs in the
thermal equilibrium state for $H_0 > 2H_{c1}$, as well as in the
critical state à la Bean for $H_0 > H^*$ \cite{BEAN}. The positive
concavity of the curves $(b)$ and $(c)$ of Fig.~2, obtained for
$H_0 < H^*$, can be qualitatively understood considering that, on
increasing the external field from $H_{c1}$ up to $H^*$, more and
more regions contribute to the mw losses.

\begin{figure}[h]
\centering
\includegraphics[width=7cm]{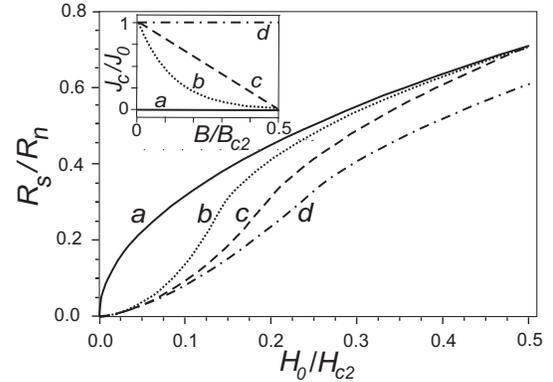}
\caption{Expected $R_s(H_0)$ curves, obtained using for the field
dependence of the critical current density the laws shown in the
inset. $T=T_c/4$, $\lambda_0/\delta_0=10^{-2}$.} \label{f3}
\end{figure}

When the field dependence of the critical current is taken into
account, the shape of the $R_s(H_0)$ curve strongly depends on the
$B$ profile due to the specific $J_c(B)$ behavior. The field
dependence of $J_c$ has been widely discussed in the literature
both theoretically \cite{KIM,IRIE,MING} and experimentally
\cite{KOBAYASHI,DESORBO,DHALLE}; different behaviors have been
suggested. In Fig.~\ref{f3} we report the expected $R_s(H_{0})$
curves, obtained considering different $J_c(B)$ laws, along with
the curve expected from the CC theory $(a)$; also in this case, we
have assumed $H_{c1}=0$. Curve $(b)$ has been obtained using
$J_c=J_{0}\exp(-B / \Gamma_0)$, curve $(c)$ with
$J_c=J_0-\alpha~B$ and curve $(d)$ with $J_c=J_0$; the inset shows
the corresponding $J_c(B)$ laws used. As one can see, the features
of the $R_s(H_0)$ curve strongly depend on the profile of $B$
inside the sample. In particular, looking at curve $(b)$, it is
evident that when the critical current verges on zero the
$R_s(H_0)$ curve approaches the one obtained from the CC theory,
as expected. This finding is expected, in any case, for magnetic
fields close to $H_{c2}(T)$, where the pinning becomes ineffective
and, consequently, $J_c\approx 0$.

It is worth noting that we have investigated the effects of the
$B$ profile determined by the field dependence of the critical
current density on the $R_s(H_0)$ curve, neglecting possible
variation of the profile due to the finite dimensions of the
superconducting sample. As it is well known, the sheet current,
induced near the sample edges, can affect the $B$ profile in the
sample \cite{brandt2}. These effects are particularly enhanced for
thin strips when the DC magnetic field is parallel to the smallest
dimension; in this case, a proper $B$ profile has to be
considered.

\section{Comparison with experiments}

In order to study the peculiarities of the $R_s(H_0)$ curve in
superconductors in the critical state, we have measured the
field-induced variations of $R_s$ in bulk Nb at low temperatures,
where the pinning effects are significant and a magnetic field
smaller enough than $H_{c2}(T)$ develops the critical state of the
fluxon lattice. The sample, of dimensions $3.3\times 2.3\times
1.3~ \mathrm{mm^3}$, has been cut from a Tokyo Denkai batch with
$\mathrm{RRR}=300$ and undergoes a superconducting transition at
$T_c \approx 9.2$~K. We have studied the Nb because the depinning
frequency \cite{GOLOSOVSKY} is small enough to suppose that the mw
current induces the fluxon lattice moving in the flux-flow regime;
furthermore, our experimental apparatus allows reaching DC
magnetic fields of the order of $H_{c2}$.

The mw surface resistance has been measured using the
cavity-perturbation technique \cite{TRUNIN}. A copper cavity, of
cylindrical shape with golden-plated walls, is tuned in the
TE$_{011}$ mode, resonating at 9.6 GHz. The cavity is placed
between the poles of an electromagnet, which generates magnetic
fields up to $\sim 1~ \mathrm{T}$. The sample is located in the
center of the cavity where the mw magnetic field is maximum. The
field geometry is that shown in Fig.~1~(b) and, therefore, the mw
current induces a tilt motion of the whole fluxon lattice. The
$R_s$ values are determined measuring the variation of the quality
factor of the cavity, induced by the sample, by an $hp$-8719D
Network Analyzer.

In order to check the presence of the critical state in the
sample, we have measured the field-induced variations of $R_s$ by
cycling the DC magnetic field from zero to the maximum value
available in our experimental apparatus and back, at different
values of the temperatures. Up to 2 K below $T_c$, $R_s$ exhibits
a magnetic hysteresis that disappears for $H_0$ higher than a
certain value, depending on the temperature. In particular, in the
range of temperatures $T \approx 2 \div 4$~K the hysteretic
behavior is detectable up to $H_0 \approx 0.4$~T; at higher
temperatures, the field range in which the hysteresis is present
shrinks; eventually, for $T > 7$~K, the $R_s(H_0)$ curve becomes
reversible in the whole range of fields investigated. These
findings confirm that, at temperatures smaller enough than $T_c$,
the external field develops a critical state in the fluxon
lattice. Furthermore, in the field range in which the hysteresis
is observed, $R_s$ does not exhibit any detectable time evolution,
from the instant in which the field value has been set, ensuring
that the critical state does not relax during the time in which
the measurements are performed. The hysteretic behavior of
microwave surface resistance will be discusses in a forthcoming
paper; in this paper, we are interested to investigate the
critical-state effects on the $R_s(H_0)$ curve at increasing
magnetic field, and to test the method we suggest for taking into
account the not uniform $B$ distribution in the sample.

Fig.~4 shows the field-induced variations of $R_s$, obtained in
the Nb sample at the two temperatures $T=4.2$~K (a) and $T=2.2$~K
(b). In the figure, $\Delta R_s(H_0,T)\equiv R_s(H_0,T)-R_{res}$,
where $R_{res}$ is the residual mw surface resistance at $T=2.2$~K
and $H_{0}=0$; the data are normalized to the maximum variation,
$\Delta R_s^{max}\equiv R_{n}-R_{res}$. $R_s$ does not show any
variation as long as $H_0$ reaches a certain value that can be
identified as the first penetration field, $H_p$. From the data,
we obtain $H_p(4.2~\mathrm{K})\approx 52$~mT and
$H_p(2.2~\mathrm{K})\approx 57$~mT; these values are smaller than
those reported in the literature for the lower critical field in
Nb superconductor because of the demagnetization effects. The
dashed and pointed lines are the curves expected from the CC
model; the continuous lines are those obtained in the framework of
our model; all the expected curves are plotted for $H_0>H_p$.
Since the depinning frequency reported in the literature for Nb
are smaller than 100~MHz \cite{GOLOSOVSKY}, we have assumed that
fluxons move in the flux-flow regime. The essential parameters to
calculate the expected curves are: $\lambda_0/\delta_0$,
$H_{c2}(T)$ and the $B$ profile inside the sample. According to
the results reported in the literature \cite{PADAMSEE,GOLUBOV}, we
have used $\lambda_0/\delta_0=3\times 10^{-2}$; however, for
$T<T_c$/2 the expected results are little sensitive to variations
of this parameter. $H_{c2}(4.2~\mathrm{K})$ has been determined
experimentally and it results 1.1~T; the value of the upper
critical field at $T=2.2$~K has been used as fitting parameter and
we have found $H_{c2}(2.2~ \mathrm{K})=1.8~ \mathrm{T}$. To take
into account the presence of the critical state, we have used a
field dependence of $J_c$ similar to that reported by Cline
\textit{et al}. \cite{CLINE} for Nb crystals; in particular, it
has been used a linear field dependence of $J_c$ at low fields,
followed by an exponential decrease for $H_0\geq H^{\prime}$. The
value of $H^{\prime}$ has been considered as fitting parameter;
the best-fit curves (continuous lines of Fig.~4) have been
obtained using $H^{\prime}(4.2~ \mathrm{K})=0.35~ \mathrm{T}$ and
$H^{\prime}(2.2~ \mathrm{K})=0.5~ \mathrm{T}$. The $B$ profiles,
obtained on increasing $H_0$, which come out from the used
$J_c(B)$ laws, are shown in the insets. The full penetration
fields calculated from the $B$ profiles are
$H^*(4.2~\mathrm{K})=0.41~\mathrm{T}$ and
$H^*(2.2~\mathrm{K})=0.56~\mathrm{T}$.

\begin{figure}[h]
\centering
\includegraphics[width=7cm]{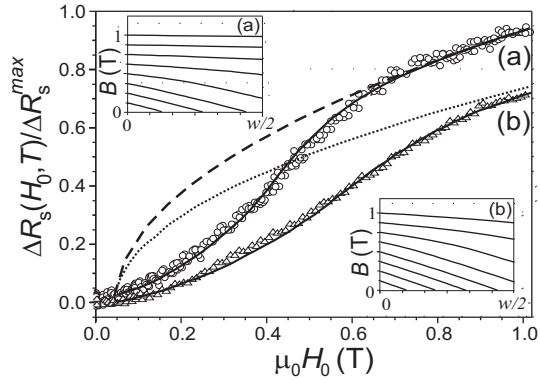}
\caption{Normalized field-induced variations of $R_s$ at: $T=4.2~
\mathrm{K}~ (a)$ and $T=2.2~ \mathrm{K}~ (b)$. Symbols are the
experimental results. Dashed and pointed lines have been obtained
from the CC model using $\lambda_0/\delta_0=3\times 10^{-2}$;
$H_{c2}(2.2~ \mathrm{K})=1.8~ \mathrm{T}$ and $H_{c2}(4.2~
\mathrm{K})=1.1~ \mathrm{T}$. The continuous lines are the
best-fit curves of the data, obtained using the same values of
$\lambda_0/\delta_0$ and $H_{c2}(T)$ and the $B$ profiles shown in
the insets.}\label{f4}
\end{figure}

As one can see, in the whole range of $H_0$ investigated the
experimental results are quite well accounted for by considering
the $B$ distribution inside the sample. At low fields, the
non-uniform $B$ profile affects to a detectable extent the energy
losses, giving rise to the positive concavity of the $R_s(H_0)$
curve. On increasing $H_0$, the slope of the field profile
decreases; the effects of the non-uniform $B$ distribution become
less and less important; eventually, when $B$ becomes roughly
uniform over the sample, the observed behavior of $R_s(H_0)$ is
consistent with that expected from the CC model.

It has been experimentally observed \cite{will}, and theoretically
justified \cite{brandt3}, that, in samples of finite dimensions,
the application of an AC magnetic field normal to the DC field
gives rise to a "shaking" of the fluxon lattice, inducing
relaxation toward the uniform distribution. The process is
particularly relevant when the amplitude of the AC field is of the
order of the full penetration field. In mw measurements with the
cavity perturbation technique, the amplitude of the mw magnetic
field is, usually, much smaller than 1 Gauss, so the process can
play a role only in very thin samples or in proximity of the
irreversibility line, where the critical current is very small. Of
course, if this process occurs, the mw field assists fluxon
distributing uniformly in the sample and, consequently, a
variation of the $R_s(H_0)$ curve is expected.

Another effect we have neglected is the deformation of the $B$
profile near the edges of the sample \cite{brandt2}; however, the
measured $R_s$ value is due to the average response of the whole
sample, and, therefore, our measurements do not allow estimating
in what extent this effect influences the experimental curves. On
the other hand, the results reported in Fig. 4 show that the $B$
profiles shown in the insets are suitable to well account for the
experimental data.
\section{Conclusion}
In summary, we have studied the field-induced energy losses in
superconductors in the critical state. We have shown that the
distribution of the local vortex magnetic field in the sample can
strongly affect the field dependence of the mw surface resistance.
The analysis has been carried out for the field geometry in which
the mw current induces a tilt motion of the fluxon lattice;
indeed, in this case, the non-uniform flux distribution most
affects the $R_s(H_0)$ curve. The expected curves have been
obtained supposing that the fluxons move in the flux-flow regime,
where the field-induced energy losses are relevant even for
applied fields much smaller than $H_{c2}$; on the other hand, the
analysis can be easily extended to a more general case, provided
that the field dependence of the depinning frequency is known. We
have highlighted that the effects of the critical state on the mw
energy losses cannot generally be disregarded to account for the
experimental data. Experimental results of field-induced
variations of $R_s$ in a Nb bulk sample at low temperatures have
been justified quite well in the framework of our model.\\

The authors are very glad to thank G. Lapis and G. Napoli for
technical assistance.

% Create the reference section using BibTeX:


\begin{thebibliography}{99}

\bibitem{gittle}J. I. Gittleman and B. Rosenblum, Phys. Rev. Lett. {\bf 16}, 734 (1966).

\bibitem{golo}M. Golosovsky, M. Tsindlekht, and D. Davidov, Supercond. Sci. Technol.
\textbf{9}, 1 (1996) and Refs. therein.

\bibitem{TALVA} J. Owliaei, S. Shridar, and J. Talvacchio, Phys. Rev. Lett. {\bf 69}, 3366
(1992).

\bibitem{noi} S. Fricano, M. Bonura, M. Li Vigni, Klinkova, and Barkovskii, Eur. Phys. J. B
{\bf 41}, 313 (2004).

\bibitem{libromw}H. Weinstock and M. Nisenoff (eds.), \textit{Microwave Superconductivity},
NATO Science Series, Series E: Applied Science - Vol. 375, Kluwer:
Dordrecht 1999.

\bibitem{CC4}M. W. Coffey and J. R. Clem, Phys. Rev. Lett. {\bf67}, 386 (1991);
Phys. Rev. B {\bf 45}, 9872 (1992); {\bf 45}, 10527 (1992).

\bibitem{BRANDT} E. H. Brandt, Phys. Rev. Lett. {\bf 67}, 2219 (1991).

\bibitem{noiPRB} A. Agliolo Gallitto, I. Ciccarello, M. Guccione, M. Li Vigni,and D. Persano
Adorno, Phys. Rev. B {\bf 56}, 5140 (1997).

\bibitem{BEAN}C. P. Bean, Phys. Rev. Lett. {\bf 8}, 250 (1962).

\bibitem{KIM}Y. B. Kim, C. F. Hempstead, and A. R. Strnad, Phys. Rev.
Lett. {\bf 9}, 306 (1962).

\bibitem{yesherun} Y. Yesherun, A. P. Malozemoff, A. Shaulov, Rev. Mod. Phys. {\bf 68},
911 (1996), and Refs. therein.

\bibitem{TINKHAM}L. Ji, M. S. Rzchowski, N. Anand, and M. Tinkham, Phys. Rev. B {\bf 47},
470 (1993).

\bibitem{SRIDHAR}Balam A. Willemsen, J. S. Derov, and S. Sridhar, Phys. Rev. B {\bf 56},
11989 (1997).

\bibitem{BS}J. Bardeen and M.~J. Stephen, Phys. Rev. {\bf 140}, A1197 (1965).

\bibitem{IRIE}F. Irie and K. Yamafuji, J. Phys. Soc. Jpn. {\bf 23}, 255 (1967).

\bibitem{MING}Ming Xu, Donglu Shi, and Ronald F. Fox, Phys. Rev. B {\bf 42}, 10773 (1990).

\bibitem{KOBAYASHI}T. Kobayashi, T. Kimura, J. Shimoyama, K. Kisho, K. Kitazawa, and
K. Yamafuji, Physica C {\bf 254}, 213 (1995).

\bibitem{DESORBO}W. DeSorbo, Phys. Rev. {\bf 134}, A1119 (1964).

\bibitem{DHALLE}M. Dhallé, P. Toulemonde, C. Beneduce, N. Musolino, M. Decroux, and R.
Flukiger, Physica C {\bf 363}, 155 (2001).

\bibitem{brandt2}E. H. Brandt, Phys. Rev. B. {\bf 54}, 4246 (1996).

\bibitem{TRUNIN}M. R. Trunin, Physics-Uspekhi {\bf 41}, 843 (1998).

\bibitem{CLINE}H. E. Cline, C. S. Tedmon, Jr., and R. M. Rose, Phys. Rev. {\bf 137}, A1767 (1965).

\bibitem{GOLOSOVSKY}M. Golosovsky, M. Tsindlekht, H. Chayet, and D. Davidov, Phys. Rev.
B {\bf 50}, 470 (1994).

\bibitem{GOLUBOV}A. A. Golubov, M. R. trunin, A. A. Zhukov, O. V. Dolgov, and S. V. Shulga,
J. Phys. I France {\bf 6}, 2275 (1996).

\bibitem{PADAMSEE}H. Padamsee, Supercond. Sci. Technol. {\bf 14}, R28 (2001).

\bibitem{will}M. Willemin, C. Rossel, J. Hofer, H. Keller, A. Erb, E. Walker,
 Phys. Rev. B. {\bf 58}, R5940 (1998).

\bibitem{brandt3}E. H. Brandt and G. P. Mikitik, Phys. Rev. Lett. {\bf 89},
027002 (2002); G. P. Mikitik and E. H. Brandt, Phys. Rev. B.
{\bf67}, 104511 (2003).

\end{thebibliography}
\end{document}